\documentclass[a4paper,num-refs]{oup-contemporary}

\journal{gigascience}

\usepackage{graphicx}
\usepackage{siunitx}
\usepackage{lmodern}

\title{Challenges in structural variant calling in low-complexity regions}

\author[1]{Qian Qin}
\author[2,3,4,\authfn{1}]{Heng Li}

\affil[1]{Brigham Women's Hospital, 75 Francis St, Boston, MA 02115, USA}
\affil[2]{Department of Biomedical Informatics, Harvard Medical School, 10 Shattuck St, Boston, MA 02215, USA}
\affil[3]{Department of Data Science, Dana-Farber Cancer Institute, 450 Brookline Ave, Boston, MA 02215, USA}
\affil[4]{Broad Insitute of MIT and Harvard, 415 Main St, Cambridge, MA 02142, USA}

\authnote{\authfn{1}hli@ds.dfci.harvard.edu}

\papercat{Data Note}

\runningauthor{Qin and Li}

\jvolume{00}
\jnumber{0}
\jyear{2025}

\begin{document}

\begin{frontmatter}
\maketitle
\begin{abstract}
\textbf{Background:}
Structural variants (SVs) are genomic differences $\ge$50 bp in length.
They remain challenging to detect even with long sequence reads, and the sources of these difficulties are not well quantified.
\vspace{0.5em}\\
\textbf{Results:}
We identified 35.4 Mb of low-complexity regions (LCRs) in GRCh38. Although
these regions cover only 1.2\% of the genome, they contain 69.1\% of confident
SVs in sample HG002. Across long-read SV callers, 77.3--91.3\% of erroneous
SV calls occur within LCRs, with error rates increasing with LCR length.
\vspace{0.5em}\\
\textbf{Conclusion:}
SVs are enriched and difficult to call in LCRs.
Special care need to be taken for calling and analyzing these variants.
\end{abstract}

\begin{keywords}
structural variant; low-complexity regions; evaluation
\end{keywords}
\end{frontmatter}


\section{Introduction}

Structural variants (SVs) are $\ge$50bp genomic variants
and may have functional impacts~\cite{Eichler:2019aa}.
Recent work based on high-quality long-read assemblies suggests
there are broadly 25,000--35,000 SVs per human individual~\cite{Liao:2023aa,Logsdon:2025ab}.
Constructed by the Genome-In-A-Bottle (GIAB) group,
the latest SV benchmark HG002-Q100 v1.1~\cite{Hansen2025.09.21.677443} contains 28,188 SVs in 2.76Gb of confident regions, consistent with the recent counts.
In contrast, published in 2020~\cite{Zook:2020aa}, the older HG002-SV benchmark v0.6 only contains 9,705 SVs in 2.66Gb.
This seems to suggest $\sim$18,000 SVs would fall in $\sim$100Mb regions if we assume the SV v0.6 regions are contained in Q100 v1.1.
Is this the correct interpretation?

This article gives the answer:
the differences between the two versions of the GIAB SV benchmarks
are primarily driven by low-complexity regions (LCRs) that harbor repeatedly occurring motifs.
The older benchmark excluded many of LCRs because it was hard to call them correctly.
Although SV callers developers have noticed the difficulties in calling SVs around LCRs~\cite{Zook:2020aa,Smolka:2024ab,Keskus:2025aa},
they have not systematically quantified the effect of LCRs in SV calling.
There is not a consensus on the number of SVs in LCRs or the error rate of them.
Here, we identified LCRs jointly from the reference genome
and the assemblies from the Human Pangenome Reference Consortium (HPRC)~\cite{Liao:2023aa},
and evaluated their impact on SV calling with multiple callers.


\section{Data Description}


We applied longdust~\cite{Li:2025aa} to GRCh38 and identified 115.4Mb of LCRs on assembled chromosomes.
We filtered about half of them that overlap with alpha and HSAT2/3 centromeric repeats found by dna-brnn~\cite{Li:2019aa}.
34.4Mb of LCRs were left when we selected LCRs of 50bp or longer.

GRCh38 only represents one human genome.
It may miss polymorphic LCRs present in other human samples but missing from GRCh38.
To look for these LCRs, we ran longdust on all 462 assemblies from HPRC
and used the results to annotate variant bubbles in the minigraph graph of these assemblies~\cite{Li:2020aa}.
A variant bubble was marked as an LCR if (a) $\ge$70\% of the sequences in the bubble were LCRs in the source assemblies, and
(b) the sequences in the bubble were not annotated as segmental duplications (SegDup) by HPRC.
Note that if an LCR falls in a long polymorphic SegDup, most of the sequences in the corresponding bubble will be annotated as SegDup but not as LCR.
This is why we put SegDup at a higher priority over LCRs during annotation.
To focus on common variants, we dropped non-GRCh38 alleles supported by $<$5 assemblies.
We ignored HG002 when counting alleles because we will use this sample for benchmarking later.

We merged the common polymorphic LCRs and GRCh38 LCRs and added 5bp to both ends of each LCR.
This resulted in a BED file with 111,067 records, covering 35.4Mb of GRCh38.
29,291 records overlap with common polymorphic LCRs in the HPRC minigraph graph.
3,918 of them are not observed on GRCh38.
16.2\% of the LCRs are intersected with the SegDup annotation from the ``genomicSuperDups'' track of the UCSC Genome Browser~\cite{Perez:2025aa}.
We see the overlap because an LCR consisting of several copies of a long repeat unit could also be considered as a SegDup.

We applied the same procedure to the T2T-CHM13 genome~\cite{Nurk:2022up}
and found 79.6Mb of LCRs, doubling the length of LCRs in GRCh38.
Most of the additional regions came from centromeric satellites that are not HSAT2/3 or alpha repeats.
If we exclude all types of satellites~\cite{Altemose:2022tv}, only 31.2Mb will remain,
comparable to the length of LCRs in GRCh38.

\section{Data Analysis}

To understand the effect of LCRs in long-read SV calling,
we measured the accuracy of SV calls stratified by LCR.
We called SVs with 11 callers and compared them to both the new HG002-Q100 v1.1~\cite{Hansen2025.09.21.677443}
and the old HG002-SV v0.6~\cite{Zook:2020aa} benchmarks.
We will also explain caveats in the truth data.

\subsection{Investigating the GIAB truth SVs}

There are 29,131 SVs of $\ge$50bp in length contained in the confident regions in the new HG002-Q100 v1.1 benchmark~\cite{Hansen2025.09.21.677443}.
943 of them have ``*'' as alternate alleles.
We manually inspected the read alignment around some of these SVs and believe they are all redundant.
Removing them from the truth left us with 28,188 SVs.
The truvari~\cite{English:2022aa} evaluation tool also filters SVs with ``*'' alleles.

The older HG002-SV v0.6 benchmark~\cite{Zook:2020aa} is only available in the GRCh37 coordinate.
To evaluate the SV calling accuracy on this benchmark,
we lifted its confident regions over to GRCh38
but we still took SVs from HG002-Q100 as the ground truth.
There are 11,985 HG002-Q100 overlapping with the lifted HG002-SV confident regions,
more than the 9,705 SVs from the older HG002-SV benchmark.
The difference is caused by the allele resolution.
Suppose both haplotypes in HG002 harbor a 6kb insertion to the same location of the reference genome.
The inserted sequences however differ by one SNP between them.
The newer HG002-Q100 benchmark
would consider this event as two heterozygous insertions,
but the older HG002-SV benchmark would merge the two insertion alleles and consider them as one homozygous insertion.
As a result, we counted 7,362 insertions in HG002-Q100 v1.1 but only 5,444 in HG002-SV v0.6, a sharp reduction.
At the same time, the allele resolution may also affect deletions.
If there are overlapping deletions of similar lengths between the two haplotypes,
HG002-Q100 will encode them two independent deletions,
but HG002-SV may merge them and thus reduce the total counts.
Overall, constructed from long-read assemblies, HG002-Q100 is more precise and more accurate than HG002-SV.

\subsection{Calling SVs from long reads}

We acquired PacBio High-Fidelity reads from HPRC~\cite{hifi-read},
aligned them to the primary assembly of GRCh38 with minimap2~\cite{Li:2018ab}
and called SVs with
cuteSV v2.1.1~\cite{Jiang:2020aa},
DeBreak v1.0.2~\cite{Chen:2023aa},
Delly v1.3.3~\cite{Rausch:2012aa},
longcallD v0.0.5~\cite{longcalld},
pbsv v2.11.0~\cite{pbsv},
Sawfish v0.12.10~\cite{Saunders:2025aa},
Severus v1.6~\cite{Keskus:2025aa},
Sniffles2 v2.6.3~\cite{Smolka:2024ab},
SVDSS v2.1.0~\cite{Denti:2023aa},
SVIM v2.0.0~\cite{Heller:2019aa}
and SVision-pro v2.4~\cite{Wang:2025aa}.
We used kanpig v1.1.0~\cite{English:2025aa} for genotyping SVs called by SVDSS as is suggested in the documentation.
We dropped SVision-pro because it did not output SV alleles and is thus incompatible with the evaluation tool.
Severus under several settings all called $\sim$40\% fewer SVs in LCRs than other callers.
We excluded it as an outlier.
Sniffles2 may optionally take tandem repeatitive regions as input,
but using this option slightly reduced its overall accuracy, so we only evaluated its default setting.

\subsection{Most SVs are located in LCRs}

\begin{figure}[tb]
\includegraphics[width=\columnwidth]{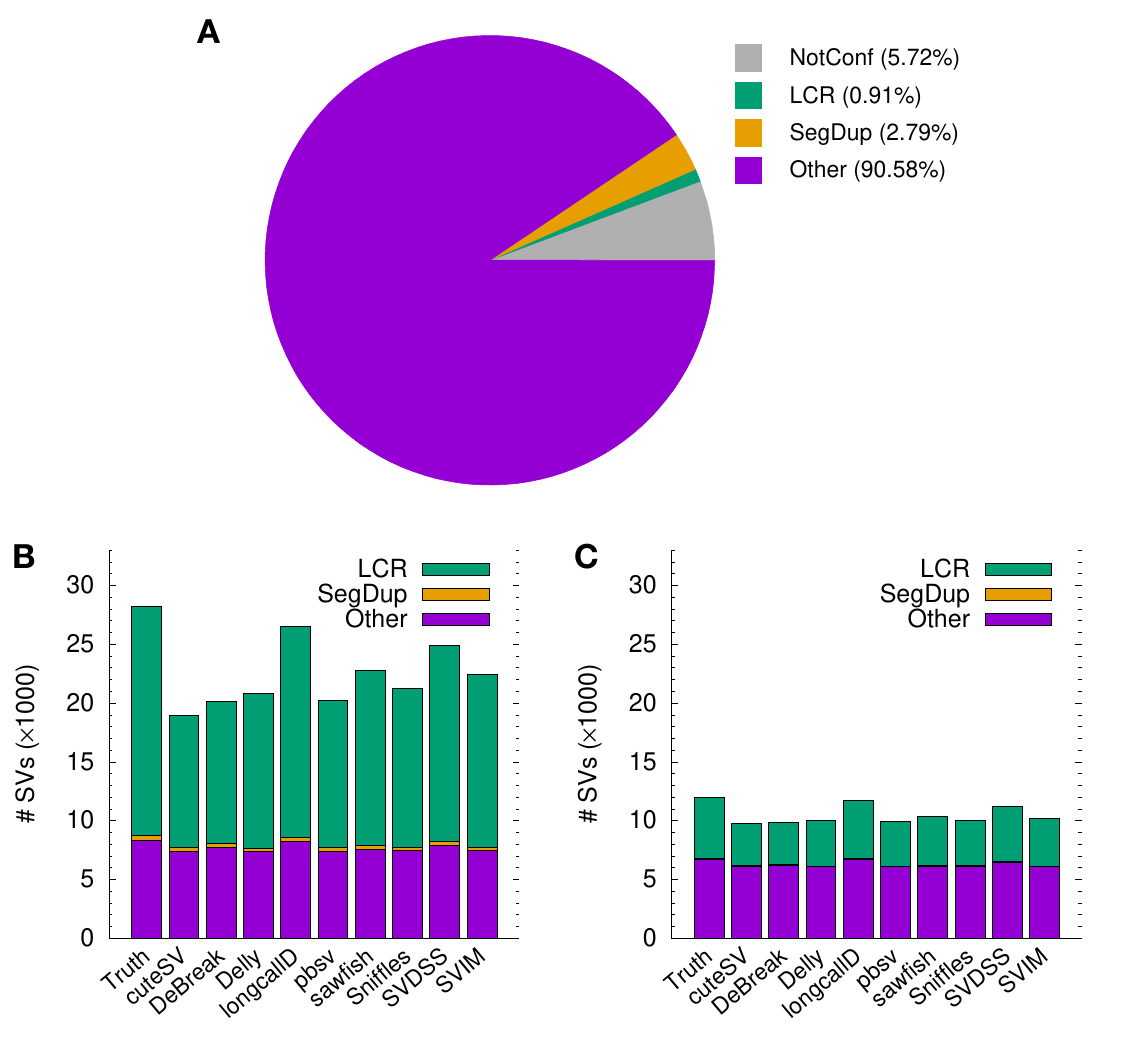}
\caption{Number of HG002 structural variants (SVs) on GRCh38.
(A) Lengths of regions.
``NotConf'' denotes not-confident regions in the HG002-Q100 v1.1 benchmark, excluding assembly gaps in GRCh38.
A region classified to a previous type will not be counted towards the next type
in the order of NotConf, LCR (low-complexity region), SegDup (segmental duplication) and Other.
(B) Number of HG002-Q100 SVs stratified by LCR, SegDup and the rest of the confident regions.
An SV is classified as LCR (or SegDup) if $\ge$70\% of its interval on GRCh38 overlaps with LCR (or SegDup).
An SV classified as LCR will not be classified as SegDup.
(C) Number of HG002-Q100 SVs in the HG002-SV v0.6 confident regions lifted over from GRCh37.}\label{fig:count}
\end{figure}

\begin{figure}[!tb]
\includegraphics[width=\columnwidth]{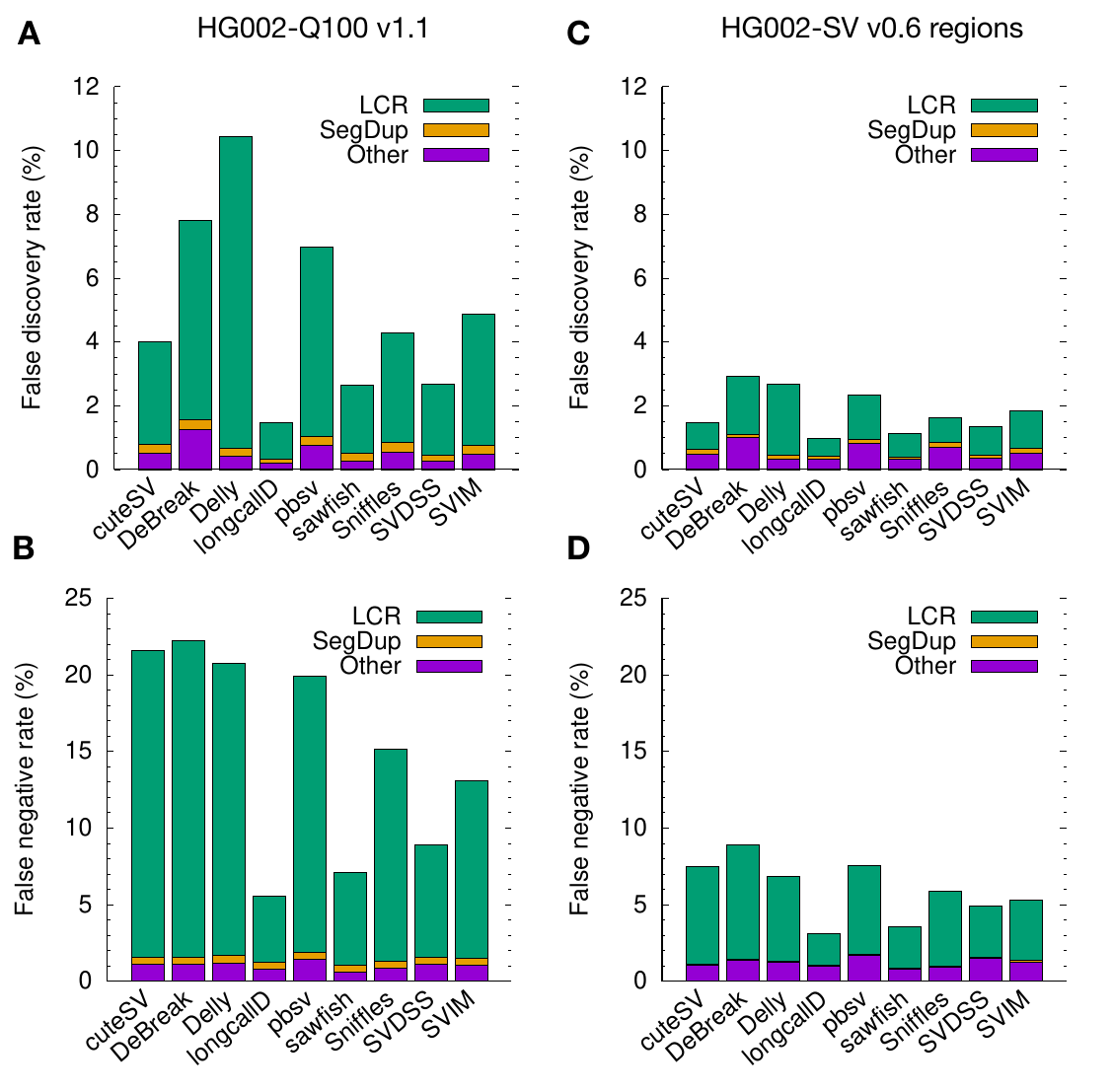}
\caption{Accuracy of SV calls.
(A) False discovery rate (FDR) of SVs in the HG002-Q100 confident regions, measured by truvari in the ``refine'' mode.
SVs are stratified to LCR, SegDup and Other in the same way as is described in Fig.~\ref{fig:count}.
(B) False negative rate (FNR) of SVs in HG002-Q100.
(C) FDR in the HG002-SV confident regions.
(D) FNR in HG002-SV.}\label{fig:acc}
\end{figure}

We stratified HG002 SVs by LCR and SegDup (Fig.~\ref{fig:count}).
For an SV to be classified as LCR or SegDup, we required it to have large overlap with LCR or SegDup regions.
Without this condition, a long deletion containing a short LCR would be falsely classified as LCR, which would inflate the number of LCR SVs.
Across the SV callers, 59.4--67.7\% of SV calls overlap with LCR, although LCR only accounts for 1.2\% of GRCh38.
SVs are highly enriched in LCR.

Whereas the numbers of ``Other'' SVs in the HG002-Q100 confident regions are similar across callers,
the numbers of LCR SVs differ greatly (Fig.~\ref{fig:count}B).
SV callers that attempt to produce haplotype-resolved SVs, such as longcallD and SVDSS, call noticeably more SVs in LCR and SegDup.
This trend is also observed in the older HG002-SV v0.6 confident regions (Fig.~\ref{fig:count}C).
In the older HG002-SV regions, there are much fewer SVs in LCR and almost none in SegDup, although the numbers of SVs
in Other regions are only reduced a little.
This indicates that the main difference between HG002-Q100 and HG002-SV comes from LCRs.

\subsection{SVs in LCRs are harder to call correctly}

We evaluated SV calls with truvari v5.3.0~\cite{English:2022aa}.
Having explored multiple truvari options,
we settled on ``{\tt bench -{}-passonly -{}-pick ac -{}-dup-to-ins}''
followed by ``{\tt refine -{}-use-original-vcfs}''
as the resulting accuracy matched our manual inspection better.
We also stratified SV calling errors by SegDup and LCR (Fig.~\ref{fig:acc}).

On the new HG002-Q100 benchmark, 31.1--39.0\% of SVs, depending on callers, are marked as ``Other'' (Fig.~\ref{fig:count}B),
but only 5.2--14.0\% of SV errors come from ``Other'' (Fig.~\ref{fig:acc}A and \ref{fig:acc}B).
This suggests SVs in the Other category are easier to call.
In contrast, the majority of errors, at 77.3--91.3\%, are located in LCRs.
SVs in SegDup are also difficult to call, but due to the small number of such SVs,
they do not contribute much to the total number of errors.

\begin{figure}[tb]
\includegraphics[width=\columnwidth]{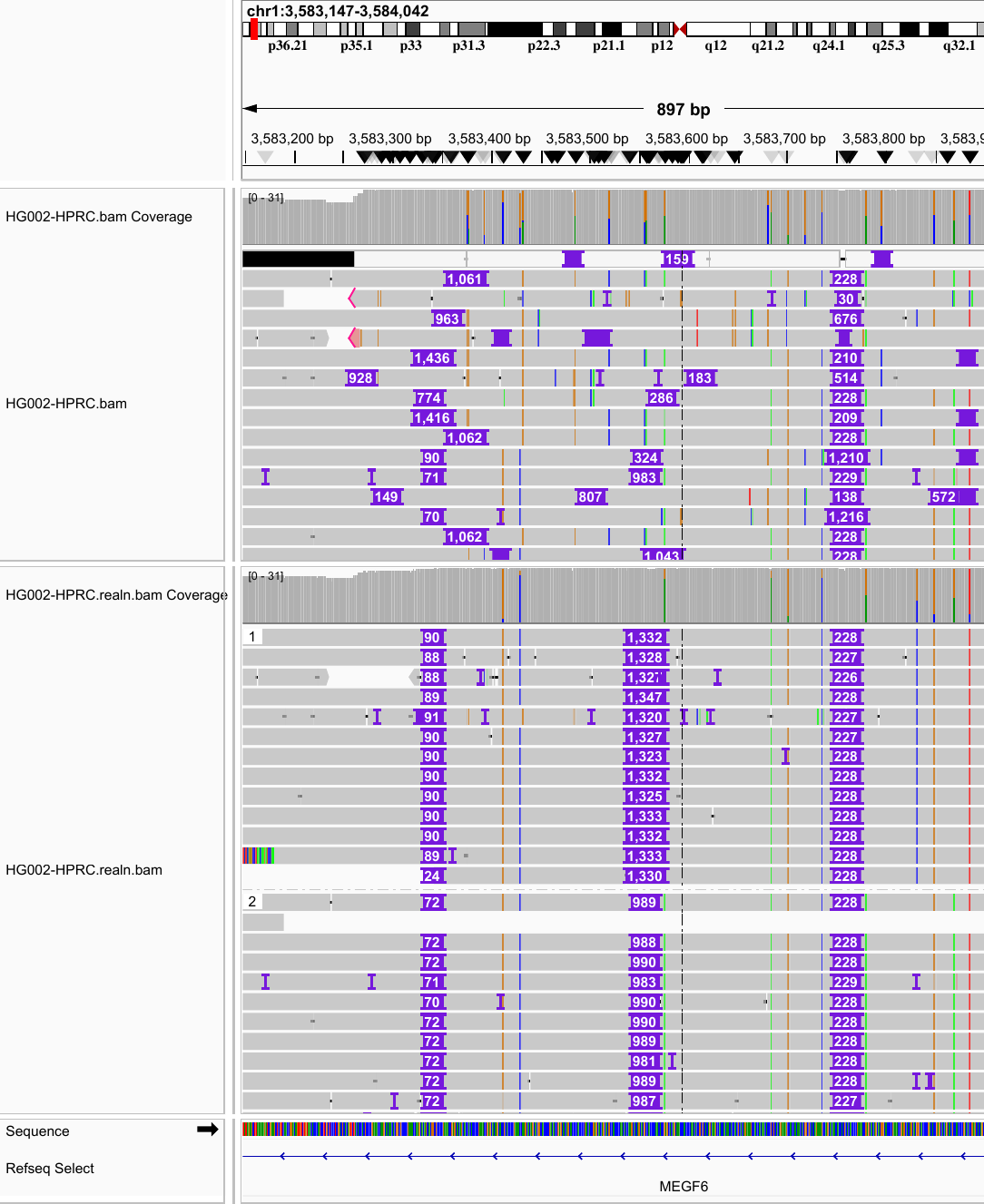}
\caption{IGV screenshot of alignment around an LCR.
The top panel shows the raw alignment by minimap2.
The bottom panel shows the phased realignment by longcallD.
There are 1650 (=90+1332+228) inserted bases on the first haplotype in total
and 1290 (=72+990+228) inserted bases on the second haplotype,
identical to the HG002-Q100 ground truth.
}\label{fig:ex}
\end{figure}

Developed in our group, longcallD achieves the lowest error rate (Fig.~\ref{fig:acc})
mainly because it performs haplotype-aware multi-sequence realignment.
As is shown in the top panel of Fig.~\ref{fig:ex}, minimap2 often produces inconsistent read alignment in long LCRs
when it does not see other reads in the same region during pairwise alignment.
It is not apparent that there are only two haplotypes in this region.
Such inconsistency would confuse most SV callers.
For this example, the SV callers in the order shown in Fig.~\ref{fig:acc}, respectively, called
+1007/+1007, +1392/+1392, +1293/+1293, +1650/+1290, +1278/+1278,
+1278/+1668, +963/+1191, +1650/+1290 and +1353/+2306 insertions on the two haplotypes.
Only longcallD and SVDSS found the precise allele lengths of +1650/+1290.
Nonetheless, truvari considered all callers correct.
The error rate of most callers would probably be higher if we required precise allele matches.


\begin{figure}[tb]
\includegraphics[width=\columnwidth]{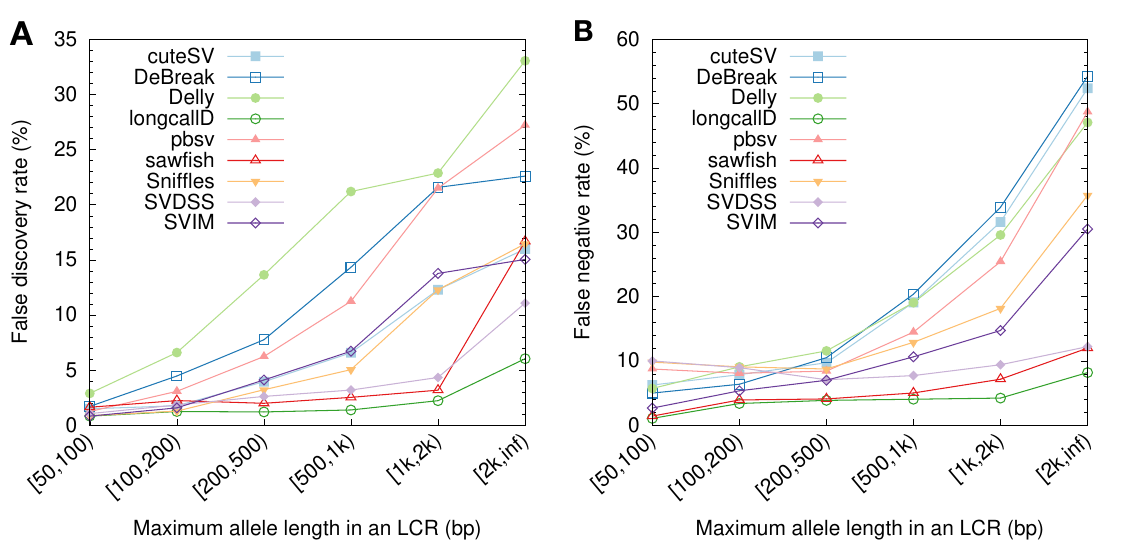}
\caption{Accuracy of SV calls stratified by the maximum allele length in LCR.
If an LCR is a common polymorphism (supported by $\ge$5 non-GRCh38 assemblies in HPRC),
the maximum allele length equals to the length of the longest allele aligned to the LCR;
otherwise, the maximum allele length equals to the length of the LCR on GRCh38.
}\label{fig:len}
\end{figure}

We further stratified the errors by the maximum allele length of each LCR (Fig.~\ref{fig:len})
and observed increased error rates with maximum allele lengths.
Some callers missed about half of SVs in $\ge$2kb LCRs,
even though HiFi reads are long enough to span most them.
Simple algorithms without realignment or reassembly
are not capable of calling SVs in long LCRs.

\section{Discussion}

LCR SVs are a distinct class.
Although LCRs only contribute to 1.2\% of GRCh38 excluding alpha and HSAT2/3 repeats,
they harbor more than half of long-read SV calls and an even higher fraction of SV calling errors.
These errors are mainly caused by inconsistent read alignment especially around long LCRs.
On the other hand, we note that LCRs may overlap with coding exons of genes that have functional impacts~\cite{Mukamel:2021aa}
and they may
also mediate gene expression~\cite{Bakhtiari:2021aa,Lu:2023aa}.
We would not want to filter all SVs overlapping LCRs.

For data analysts, we recommend stratifying SVs by LCR
as LCR SVs are enriched with errors and are resulted from different biological processes.
For developers, we would like to emphasize the critical role of realignment or local reassembly in accurate SV calling.
Most SVs in LCRs can still be called to decent accuracy with good algorithms.

Given accurate long reads at high coverage, we may also assemble
the reads with haplotype-resolved assemblers~\cite{Cheng:2021aa,Rautiainen:2023ab}
and call variants from assembly-to-reference alignment~\cite{Li:2018aa}.
Performing phasing and alignment within each haplotype,
these assemblers are more powerful than most SV callers.
As a matter of fact, the HG002-Q100 truth was derived this way.

We have only analyzed one human sample in this article.
If mainstream SV callers are already struggling with long LCRs,
merging their calls across different samples will be more problematic.
When haplotype-resolved assembly is possible,
calling variants across samples with pangenome-based methods~\cite{Li:2020aa,Hickey:2024aa,Garrison:2024ab}
will be the preferred approach as conducting multi-sequence alignment across samples,
such methods can produce more consistent SV representations.
They may also struggle with highly variable LCRs,
but will do better than traditional SV merging in most cases.

\section{Data Availability}

LCRs are available at \url{https://doi.org/10.5281/zenodo.10903864}
(file ``{\tt chm13v2.lcr-v4.bed.gz}'' and ``{\tt hg38.lcr-v4.bed.gz}'').
Scripts used for producing the LCRs and plots
can be found at \url{https://github.com/lh3/lcr-sv}.

\section{Declarations}

\subsection{List of abbreviations}

GIAB: Genome-In-A-Bottle group;
HPRC: Human Pangenome Reference Consortium;
kb: kilobase;
LCR: low-complexity regions;
Mb: megabase;
SegDup: segmental duplication;
SV: structural variant.

\subsection{Competing Interests}

The authors declare they have no competing interests.

\subsection{Funding}

This work is supported by National Institute of Health grant
R01HG010040,
R01HG014175,
U24CA294203,
U01HG013748
and U41HG010972 (to H.L.).

\subsection{Author's Contributions}

H.L. conceived the project.
Q.Q. produced structural variant calls.
Q.Q. and H.L. analyzed the data and drafted the manuscript.

\section{Acknowledgements}

We would like to acknowledge the National Genome Research Institute (NHGRI) for
funding the following grants supporting the creation of the human pangenome
reference: U41HG010972, U01HG010971, U01HG013760, U01HG013755, U01HG013748,
U01HG013744, R01HG011274, and the Human Pangenome Reference Consortium
(BioProject ID: PRJNA730823).

\bibliography{lcr-sv}

\end{document}